\documentclass[12pt]{article}

\newcommand{\no}{\noindent}

\newcommand{\be}{\begin{equation}}

\newcommand{\ee}{\end{equation}}

\newcommand{\bea}{\begin{eqnarray}}
\newcommand{\eea}{\end{eqnarray}}
\newcommand{\sss}{{\cal S}}
\newcommand{\ores}{O_{res}({\cal H})}
\newcommand{\ures}{U_{res}({\cal H})}

\usepackage[psamsfonts]{amsfonts}

\begin{document}

\vskip .5cm
\huge
\centerline{On Baryon Number Non-Conservation}
\centerline{in Two-Dimensional O(2N+1) QCD}

\vskip 2cm
\Large
\centerline{Tamar Friedmann$^*$}

\vskip .5cm
\normalsize
\centerline{\it Massachusetts Institute of Technology}
\centerline{\it Cambridge, MA}
\vskip 3cm
\renewcommand{\baselinestretch}{1.5}
\abstract{}

We construct a classical dynamical system whose phase space is a certain 
infinite-dimensional Grassmannian manifold, and propose that it is equivalent to the large $N$
limit of two-dimensional QCD with an $O(2N+1)$ gauge group. In this theory, we find that
baryon number is a topological quantity that is conserved only modulo 2.
We also relate this theory to the master field approach to matrix models.

\vskip 2cm
\no Keywords: Two-dimensional QCD; classical dynamical systems; topological solitons; baryon number; Grassmannian manifold; conservation law; 't Hooft equation for mesons; matrix model; master field.

\vfill 

\no $^*$E-mail: tamarf at mit.edu \hfill Accepted for publication at {\it Int. J. Mod. Phys. A}.

\thispagestyle{empty}
\newpage
\setcounter{page}{1}

\renewcommand{\baselinestretch}{1.5} 

\normalsize

\section{Introduction}An interesting type of a theory of the strong 
interactions is one in which there is a topological quantity which can 
be related to hadrons; several theories relate such a quantity to baryon 
number. The plausibility of such theories  may be enhanced by their being 
consistent with the well-accepted gauge theory of the strong 
interactions - i.e. with QCD.

One paper by Rajeev proposes such a theory in two 
dimensions, and shows its equivalence to the large $N$ limit of $SU(N)$ QCD \cite{Rajeev:1994tr}. The theory is a classical one
 in which the phase space is an infinite dimensional Grassmannian 
manifold, and baryon number is the topological invariant that 
corresponds to the ${\bf Z}$-many connected components of the manifold.  Rajeev's 
theory is tied into the gauge theory of the strong 
interactions via the equation for the mass of the mesons, which agrees with the one
derived by 't Hooft in the large $N$ limit of $SU(N)$ gauge theory \cite{'tHooft:1974hx}.

Here, we propose an analog of Rajeev's theory for the 
large $N$ limit of $O(2N+1)$ QCD in two dimensions\footnote{A different analog was considered in \cite{Toprak:2002kx, Toprak:2002kw}.}, and arrive at an interesting consequence
regarding baryon number non-conservation. For our phase space, we construct a different 
infinite dimensional Grassmannian manifold for which the meson equation
is the same as the one arrived at in the $O(2N+1)$ gauge theory. Our phase space has a topological
invariant of its own, which also corresponds to baryon number. Unlike the $SU(N)$ case, in this case
there are only two connected components; particles with even baryon 
number are assigned to one component, and those with odd baryon number 
to the other 
component. We thereby discover that baryon number is conserved only modulo 
two, and baryons may annihilate or be created in pairs. This fact, as we will explain, is 
independently true of QCD with an $O(2N+1)$ 
gauge group.

We also show that our theory is related to the master field approach to large N matrix models via a master field whose 
commutation relations match those of our theory.

This paper is organized as follows. In Section \ref{tH}  we review  't Hooft's derivation of the equation for mesons in two dimensions 
in the context of the 
large $N$ limit of $U(N)$ QCD. In Section \ref{sun},  we review
the theory developed by Rajeev, concentrating on the derivation of the meson equation and the topological invariant corresponding to baryons. 
In Section \ref{on} we present our analog for QCD with an orthogonal gauge group along with the topological invariant corresponding to baryons. In Section 
\ref{bnnc} we show that our result about baryon number non-conservation agrees with two-dimensional QCD  with an odd orthogonal gauge group. 
In Section \ref{matrix} we discuss the relation between our model and the master field approach to matrix models for large $N$ QCD. 
We conclude in Section \ref{next} with
several suggestions for further work.


\section{Planar diagrams and mesons in two dimensions}\label{tH}
The planar diagram theory developed by 't Hooft \cite{'tHooft:1973jz} and elaborated upon by 
Witten \cite{Witten:1979kh} provided a simplification of QCD gauge 
theory. 't Hooft considered QCD with color gauge group $SU(N)$ in the limit 
$N\rightarrow \infty $ with $g^2N$ held fixed, and arrived at the notion that in this limit, 
 only planar diagrams need to be considered, all others being 
suppressed by factors of 1/N. Still, however, calculations remain 
complicated even when they include only the planar diagrams.

For the case of two-dimensional QCD, 't Hooft showed that 
a further simplification arises, and derived an equation for the meson 
spectrum \cite{'tHooft:1974hx}. 
The derivation is briefly reviewed as follows: 

We start with the QCD lagrangian 
\be \label{L} {\cal L}= \frac{1}{4} F_{\mu \nu}F^{\mu \nu} - \overline{\psi}  
(i\gamma ^\mu D _\mu +m)\psi \ee
where $F_{\mu \nu}$ is the field strength, $D_\mu $ the covariant 
derivative, $\psi $ the quark wave function, and $m$ the quark mass.
 In two dimensions, the index $\mu$ runs over $\{ 0,1 \}$. Switching to 
light-cone coordinates, where $x^{\pm}=(x^1 \pm x^0)/\sqrt{2}$, 
$p_\pm =(p_1\pm p_0)/\sqrt{2}$, $A_\pm =(A_1\pm A_0)/\sqrt{2}$, and $g_{ab}
=\delta _{ab}-1$, we impose the gauge $A_- =A^+ =0$ and the lagrangian simplifies
to:
\be {\cal L}= -\frac{1}{2} Tr(\partial _- A_+)^2 - \overline{\psi}  
(i\gamma ^\mu \partial _\mu +m+g\gamma _- A_+)\psi .\ee
There is only one vertex, the gauge fields do not interact with themselves, and the feynman rules simplify considerably. 

To derive the equation for the meson, we first need the dressed propagator, $G(k,m)$, for the quark. Let $i\Gamma (k)$ be
the amplitude for the (planar) irreducible quark self-energy diagram. Then 
$G(k,m)$ is a sum of diagrams created from the
irreducible self-energy blob, and is equal to
\be \label{Gk} G(k, m)=\frac{-ik_- }{ m^2+2k_+ k_- -k_- \Gamma (k) -i\epsilon } .
\ee
An expression for $\Gamma (k)$ can be derived from a bootstrap equation
from which a logarithmic 
UV divergence is removed and into which an IR cutoff $\lambda $ is 
introduced. The expression is
\[ \Gamma (k)=\Gamma (k_-)=-\frac{g^2}{\pi}\left ( \frac{sgn(k_- )}
{\lambda}-\frac{1}{p_- }\right ) ,\] 
and using it in equation (\ref{Gk}) gives 
\begin{eqnarray}\label{G} G(k,m)&=& \frac{-ik_-}{m^2+2k_+ k_- -g^2 / \pi +g^2 
|k_-|/ \pi \lambda -i\epsilon }\nonumber \\ 
&=&  \frac{-ik_-}{M^2+2k_+ k_-  +g^2 |k_-|/ \pi \lambda -i\epsilon },
\end{eqnarray}
where $M^2= m^2-g^2/\pi$. Since $\lambda $ is small, we see that 
the poles of $G(k)$ occur at $k_+\rightarrow \infty $, which means there 
is no physical single quark state - no free quarks. 

To find the spectrum for mesons, we consider a blob out of  
which come a quark and an antiquark. 
 
One can derive a bootstrap equation for such a blob. Let $\psi (p,r)$ represent 
a blob out of which go a quark with mass $m_1$ and momentum $p$ and an 
antiquark of mass $m_2$ and momentum $p-r$.
 Then a bootstrap equation for it is given by
\small
\be \psi (p,r)= -\int \frac{d^2 k}{(2\pi )^2 i}4g^2 G(p-r, m_2) 
G(p, m_1) \frac{1}{k_- ^2} \psi (p+k, r) . \ee
\normalsize
Defining $ \phi (p_-,r)=\int dp_+ \psi (p_+, p_-, r)$, and substituting 
the expressions for $G(k,m)$ from equation (\ref{G}), we get after some algebra
\be \label{integ} \mu ^2 \phi (x)= \left ( \frac{\alpha _1}{x}+\frac{\alpha _2}{1-x} 
\right )\phi (x) -P\int _0^1  \frac{\phi (y)}{(y-x)^2} dy , 
\ee 
\no where $\mu ^2  \pi / g^2 = r\cdot r =-2r_+ r_-$ (i.e. $\mu $ is the 
meson mass in units of $g/ \sqrt{\pi}$), $\alpha _i=  \pi m^2_i /g^2 -1$, 
$x= p_-/ r_-$, ``P'' stands for the principal value integral, i.e.
\[ P\int  \frac{\phi (y)}{(y-x)^2} dy = \frac{1}{2} \int 
\frac{\phi (y+i\epsilon )}{(y+i\epsilon -x)^2}dy + \frac{1}{2} \int 
\frac{\phi (y-i\epsilon )}{(y-i\epsilon -x)^2}dy ,\]
which is finite, and where the infrared cutoff $\lambda$ has disappeared. Equation (\ref{integ}) is known as 't Hooft's integral 
equation for mesons. The spectrum (i.e. the eigenvalues $\mu ^2$) can be 
shown to be a discrete set of positive eigenvalues if $m_i ^2>0$.

When we replace $U(N)$ with $O(M)$ as the gauge
group (where $M$ may be even or odd), the derivation of the integral equation is the same so the same equation holds for mesons in two-dimensional 
QCD with an $O(M)$ gauge group. 


\section{$SU(N)$ QCD on a Grassmannian Manifold}\label{sun}
Now we review the main ingredients of Rajeev's construction of the large N
limit of two-dimensional QCD as a classical dynamical system \cite{Rajeev:1994tr}.

There are three necessary ingredients for a classical dynamical system:
a phase space, a symplectic form, and a Hamiltonian. Here, the phase space 
is given by an infinite dimensional Grassmanian manifold, described in two 
equivalent ways. The first description is the following set of operators 
on a polarized Hilbert space ${\cal H}={\cal H}_+ \oplus {\cal H}_-$:
\be \label{gr11} Gr_1=\{ \Phi \; | \;  \Phi ^\dagger = \Phi \;\; ; \; \Phi ^2 = 1 \;
\; ;\; \; [\epsilon , \Phi ] \mbox{ is Hilbert-Schmidt} \},
\ee 
\no where $\epsilon =\pm 1$ on ${\cal H}_{\pm}$, and an operator 
$A$ is Hilbert-Schmidt if $tr(A^TA)< \infty $. The Hilbert space is taken 
to be the space of square integrable complex valued functions on the real line (or the circle, 
with radius taken to infinity); it is spanned by Fourier modes: 
$F(\theta )=\sum _{-\infty}^{\infty} F_m e^{im\theta }$; ${\cal H}_+$ 
is defined as the span of $\{ e^{im\theta }, \, m\geq 0 \}$, and 
${\cal H}_-$ is its orthogonal complement.

It will be convenient at times to rewrite this description in terms  of $M=\Phi - \epsilon $:
\be \label{Gr1M} Gr_1 = \{ M \; | \;  M ^\dagger = M \;\; ; \; 
M^2+\epsilon M +M\epsilon
 = 0 \; \; ;\; \; [\epsilon , M ] \mbox{ is Hilbert-Schmidt} \} .
\ee
The second description of the space, equation (\ref{Gr1M}), comes from the following realization:
 each $\Phi \in {Gr_1}$ can be 
diagonalized to $\epsilon $ via the action of a unitary transformation $g$ 
on the same Hilbert space: 
$g\phi g ^{\dagger}=\epsilon$, where by ``unitary'' we mean that $g$ 
is an element in $U_{res}({\cal H})$ defined by 
\be U_{res}({\cal H})= \{ g \, | \, g^{\dagger}g=1; [\epsilon , g] \; \; 
\mbox{is Hilbert-Schmidt} \} \; .\ee 
\vskip 2cm
The stabilizer of $\epsilon $ under 
this action of the unitary group is $ U({\cal H}_+) \times U({\cal H}_-)$. 
Therefore, we may view $Gr_1$ as the coset space 
\be \label{gr12} {Gr_1}=\frac{U_{res}({\cal H})}{U({\cal H}_+) \times U({\cal H}_-)}\; ,
\ee 
which can be recognized as an infinite dimensional analog of finite
dimensional Grassmannian spaces such as $U(n)/[U(r)\times U(n-r)]$. 

To derive equations of motion, we need a symplectic form - which would give 
the Poisson Brackets - and a Hamiltonian. The symplectic form on this space 
is given by 
\be \label{sp} \omega (U, V)= -\frac{i}{8}Tr \Phi [U(\Phi),V(\Phi)] \; , \ee
where $U,V$ are tangent vector fields to $Gr_1$ (they, too, are 
hermitian operators, and they satisfy $V(\Phi)\Phi + \Phi V(\Phi )=0$).

While we will not be using this symplectic form for our case, we review here some of its properties: in addition to $\omega(U,V)$  being a closed, non-degenerate two-form, it 
is invariant under the unitary group action, $\Phi \rightarrow g\Phi g^\dagger$
or $M\rightarrow gMg^\dagger + g[\epsilon, g^\dagger ]$. For the infinitesimal
form of this action, $\Phi \rightarrow \Phi +V_u= 
\Phi +i[u,\epsilon +M]$ where $e^{iu}=g$ and $u$ is Hermitian, there is a function $f_u$ associated
with each $V_u$ defined by 
\[ \omega (V_u, \cdot)=df_u\cdot \]
namely
\[ f_u(M)=-\frac{1}{2} TrM u \, .\]

We note here that while in a finite dimensional space, the associated 
function $f_u$ of any vector field 
always exists (the symplectic form, which is nondegenerate, is "invertible"), this is not always the case 
for an infinite dimensional
space, but such a function does exist here.

The Poisson Brackets of two such functions are defined in terms of $\omega$ by
\[ \{ f_u,f_v\} = \omega(V_u, V_v) \, .\]
We get the following relation:
\be \label{pb} \{ f_u,f_v \}(M) = f_{i[u,v]}(M)-\frac{i}{2}Tr
[\epsilon , u]v\; . \; \; 
\ee 
We will now translate this into integral kernel language: the integral 
kernel $M(x,y)$ of the operator $M$, (also known as a master field - see Section \ref{matrix}), for any $F\in {\cal H}$, is defined by
\be \label{kernel} (MF)(x)= \int M(x,y)F(y)dy \; . \ee
The kernel $\epsilon (x,y) =\epsilon (x-y)$ for the 
operator $\epsilon$ is defined similarly (and is known as the Hilbert transform operator). 
Furthermore, a trace such as $TrMu$ is rewritten as $\int dx \, dy \, 
M(x,y)u(y,x)$. Now we can rewrite equation (\ref{pb}) in terms of integral 
kernels: 
\be \label{pbM} \frac{i}{2}\{M(x,y) ,M(z,w) \}= 
\delta (x-w) [M+\epsilon](z,y) -\delta (y-z)
[M+\epsilon](x,w)
\ee
(one can go back from this form to equation (\ref{pb}) by multiplying by 
$u(y,x)v(w,z)$ and integrating over all four variables.)
The Hamiltonian is taken to be
\be \label{hami} H(M)=\int dxdy h(x-y)M(x,y) -\frac{1}{2} g^2\int dxdyG(x-y)M(x,y)M(y,x)~ ,
\ee 
where the first term is the kinetic term, the second is the potential 
(interaction) term, and $g$ is a constant parameter. The kernels $h(x-y)$ and $G(x-y)$ are given by  
\[ h(x-y)=\frac{i}{2}(-\delta '(x-y) + \frac{i}{2}sgn(x-y)) ~ ,\hskip 1.5cm 
G(x-y)=-\frac{1}{2}|x-y| ~ ,\]
 which are the kernels of the Fourier transforms of
\[ h(p)=\frac{1}{2}(p+\mu ^2 /p) ~ ,\hskip 1cm G(p)=\frac {1}{p^2} ~ .\]
The kinetic term above is derived  using 2-dimensional space with 
quasi-light cone coordinates: let $u=x^0-x^1$, $x=x^1$, with metric 
$ds^2=du(du+2dx)$. The momenta in these coordinates are $p_u=p_0$ (associated
with energy) and 
$p=p_x =p_0+p_1$. In these coordinates, the invariant mass $\mu ^2$ is given by 

\[ 2pp_u-p^2=\mu ^2 ~ ,\] 
so 
\[ p_u=\frac{1}{2}(p+\frac {\mu ^2}{p}) ~ ,\]
where $p_u$ can be interpreted as the kinetic energy.

Now, the equation of motion, with $t$ as the time variable, is
\small
\begin{eqnarray} \frac{i}{2}\frac{\partial M(x,y, t)}{\partial t} &=&
\frac{i}{2}\{H,M(x,y) \} 
\nonumber \\ &=& \int dz \, [h(x-z)M(z,y)-M(x,z)h(z-y)] \nonumber \\ 
&&  + g^2\int dz \, G(y-z)\epsilon (x,z) M(z,y)-G(z-x)\epsilon (z,y)M(x,z) 
\nonumber \\ &&  +g^2 \int dz \, M(x,z)M(z,y)[G(y-z)-G(z-x)]~ . \nonumber
\end{eqnarray} \normalsize
To arrive at the meson equation, 
we take a linear approximation around the 
vacuum $\Phi = \epsilon$ ($M=0$), so we neglect the terms quadratic in M:
\small
\begin{eqnarray} \frac{i}{2}\frac{\partial M(x,y, t)}{\partial t} &=&
\frac{i}{2}\{H,M(x,y) \} 
\nonumber \\ &=& \int dz \, [h(x-z)M(z,y)-M(x,z)h(z-y)] \nonumber \\ 
&&  + g^2\int dz \, G(y-z)\epsilon (x,z) M(z,y)-G(z-x)\epsilon (z,y)M(x,z)~ .
\end{eqnarray}
\normalsize
Translating to momentum space, where 
\be M(p,q)=\int dx\, dy\, e^{i(-px+qy)}M(x,y) ~ , \ee
the linearized equation becomes 
\small
\be \frac{i}{2}\frac {\partial M(p,q, t)}{\partial t}= [h(p)-h(q)]M(p,q) \; +\;  g^2 
(sgn(p)-sgn(q))\int \frac{dr}{r^2}M(p-r, q-r) ~ .\ee
\normalsize

Now, define
\be \label{cov} P=p-q ~ ,\hskip 1cm  \xi = p/P ~ ,\hskip 1cm \chi (P, \xi)= PM(P\xi, 
-(1-\xi )P) ~ .\ee
The quantity $P$ has the meaning of momentum, since the translation 
$M(x,y)\rightarrow M(x+a, y+a)$ yields in momentum representation 
$M(p,q)\rightarrow e^{i(-pa+qa)}M(p,q)$. Furthermore, the 
constraint (see equation (\ref{Gr1M})) $M^2+M\epsilon +\epsilon M=0$, 
which to 
first order in $M$ is $M\epsilon +\epsilon M=0$ or in kernel language 
$\int dy \, M(x,y)\epsilon (y,z) + \epsilon (x,y)M(y,z) =0$, has the Fourier
transform
\[ (sgn(p)+sgn(q))M(p,q)=0 \, .\]
We also have $M^\dagger =M$, or $M^* (y,x)=M(x,y)$, which
translates to $M^*(p,q)=M(q,p)$ in momentum space. Putting these together 
we note that $M(p,q)$ may 
be non-zero only when the signs of $p$ and $q$ are opposite, and due to 
the hermiticity condition, it is sufficient to consider the case where 
$p>0, q<0$. These conditions also imply that $\xi$ of equation (\ref{cov}) 
ranges between 0 and 1. 
Now, changing from $p,q, M(p,q)$ to $P, \xi, \chi (P, \xi)$, and using the 
ansatz $M(p,q,t)=e^{-ip_u u}M(p,q)$ we get
\small
\be (2p_u P-P^2)\chi (\xi)={\cal M}^2 \chi (\xi)= (\frac{\mu ^2}{\xi}+ 
\frac{\mu ^2}{1-\xi})\chi (\xi)+4g^2 \int_0^1 \frac{d\xi '}{(\xi -\xi ')^2} 
\chi(\xi ') \; ,\ee
\normalsize
which is 't Hooft's integral equation for mesons (see equation (\ref{integ})).

Now we turn to baryons, which are topological solitons in this 
classical theory, and discuss the relevant topological properties of 
$Gr_1$: for an element 
$\Phi \in {Gr_1}$, let $V^+_{\Phi}$ be the subspace of ${\cal H}$ 
which is the $+1$ eigenspace of $\Phi$, and let $V^-_{\Phi}$ be the subspace
of ${\cal H}$ 
which is the $-1$ eigenspace of $\Phi$ (the condition 
$\Phi ^2 = 1$ leads to the fact that $\Phi$ has eigenvalues $\lambda 
= \pm 1$); also, let $M_\Phi = \Phi -\epsilon$ as before. The fact that $[\epsilon ,
\Phi]$ is
Hilbert-Schmidt implies (see \cite{Pressley:1988qk}) that the number of independent 
vectors in ${\cal H}$
which have eigenvalue $+1$ under the action of $\epsilon$ but become $-1$
eigenvectors under $\Phi$ is finite, i.e. $dim(V^-_\Phi \cap {\cal H}_+)$
is finite; similarly, $dim(V^+_\Phi \cap {\cal H}_-)$ is finite. Define 
\be \label{rank}  J(\Phi)=I_+(\Phi )-I_-(\Phi)=dim(V^-_\Phi \cap {\cal H}_+) 
- dim(V^+_\Phi \cap {\cal H}_-)\; , \ee
\no which can also be written 
\begin{eqnarray}  \label{J} J(\Phi)&=& \left( -\frac{1}{2}Tr (\Phi - 
\epsilon) | _{{\cal H}_+}\right) - \left( \frac{1}{2}Tr (\Phi - \epsilon) 
| _{{\cal H}_-}\right)  \nonumber \\ &=&  -\frac{1}{2}Tr M_\Phi = -\frac{1}{2}
\int dx M(x,x) \; .
\end{eqnarray}
\no The quantity $J(\Phi)$ can take on any integer value and is known as 
the virtual rank, or index, of the operator $\Phi$. It is a topological 
invariant: smoothly varying $\Phi $ leaves the index fixed. This index 
divides $Gr_1$ into ${\bf Z}$ connected components. 

The integer $J(\Phi )$ 
corresponding to each $\Phi $ is the topological 
invariant which corresponds to the baryon number.


\section{$O(2N+1)$ QCD on a(nother) Grassmannian}\label{on}

Here we construct a manifold ${\cal S}$, a real analog of $Gr_1$, as the 
phase space for a classical dynamical system, which we propose to be
equivalent to the large $N$ limit of $O(2N+1)$ QCD. The hamiltonian will be taken
to be the analog of the one in Section \ref{sun}; 
as for the symplectic form, we take it to be the canonical one on the loop group
 $LO_{2n}$, which is closely related to ${\cal S}$. We will show that mesons in this theory satisfy 
't Hooft's integral equation. In the next section 
we will discuss baryons in this theory.

As in Section \ref{sun}, the phase space can be described in two ways. 
The definition analogous to equation (\ref{gr11}) is 
\be \label{sphi} {\cal S}=\{ \Phi \; | \;  \Phi ^T = \Phi \;\; ; \; \Phi ^2 = 1 \;\; ; 
     \; \; [\epsilon , \Phi ] \mbox{ is Hilbert-Schmidt} \}~,  
\ee
where here $\Phi $ are operators with real matrix elements, and we use
the transpose instead of hermitian conjugate. Again, there is a description
in terms of $M=\Phi -\epsilon $:
\be {\cal S}=\{ M\, |\, M^T=M\, ; \;\; M^2+\epsilon M +M\epsilon =0; \;\;
[\epsilon, M] \mbox{ is Hilbert-Schmidt} \} ~ .\ee
Following a reasoning similar to that in Section \ref{sun}, we get the analog of equation 
(\ref{gr12}):
\be \label{sores}{\cal S}=\frac{O_{res}({\cal H})}{O({\cal H}_+) \times O({\cal H}_-) }~ ,
\ee
where \be O_{res}({\cal H})=
\{ g \, | \, g^Tg=1; [\epsilon , g] \; \; \mbox{ is Hilbert-Schmidt} \} ~ .\ee

Turning to the symplectic form, we note that for ${\cal S}$, the  
form analogous to the one defined in equation (\ref{sp}) is not 
nondegenerate\footnote{Proving 
nondegeneracy for the form $\omega$ in equation (\ref{sp}) (see \cite{Rajeev:1994tr}) involves looking at $\omega $ 
 at the chosen point 
$\Phi =\epsilon $, where we have $\omega _{\epsilon }(U,V) \propto tr(u^\dagger v 
-v^\dagger u)$ for appropriate $u,v$. At this chosen point, $\omega _{\epsilon }$ 
is non-degenerate, but for ${\cal S}$, we would have $tr(u^T v -v^T u)$ which 
is identically zero. Unlike here, in \cite{Toprak:2002kw}, a complexified version of equation (\ref{sores}) was used, and their analog of equation (\ref{sp}) is still non-degenerate.
}. Furthermore, the Poisson Brackets of equation 
(\ref{pbM}) are inconsistent with the constraint $\Phi ^T = \Phi$, which 
translates into $M(x,y)=M(y,x)$. Therefore, we shall describe a different space, 
the loop 
group $LO_{2n}$, which is closely related to ${\cal S}$ and on which there 
is a canonical symplectic form.

The group $L O_{2n}$, which is the loop group of the orthogonal 
group $O(2n)$, 
is defined as the group of smooth maps from the circle to a 
closed curve in $O(2n)$:

\be L O_{2n} =\{ \gamma :S^1\longrightarrow O(2n) \} ~ ,\ee
where the group multiplication is given by pointwise multiplication in $O(2n)$.
 Since the fundamental group of $O(2n)$ is ${\bf Z}_2$:

\be \label{fund} \pi _1 (O(2n))={\bf Z}_2 ~ ,\ee
 
\no we see by definition that $L O_{2n}$ has two connected components. 
We shall use them  
below in connection with baryon number.

The group $LO_{2n}$ is related to ${\cal S}$ because it is 
embedded in $O_{res}({\cal H})$: let ${\cal H}$  
be the Hilbert space of square integrable functions from the circle to 
${\bf R}^{2n}$.\footnote{Note that this Hilbert space, $L^2(S^1; {\bf R}^{2n})$, may be identified with the Hilbert space $L^2 ({\bf R} ; {\bf C})$  defined as ${\cal H}$ at the beginning of Section 3. See \cite{Pressley:1988qk}.}  Then $\gamma \in LO_{2n}$ acts on $F\in {\cal H}$ by pointwise 
multiplication given by an operator $M_\gamma$:
\be \label{corresp} (M_\gamma \cdot F) (\theta)=\gamma (\theta)F(\theta) ~.\ee
If we write $\gamma (\theta)=\sum \gamma _k e^{ik\theta}$, $\gamma _k \in O(2n)$, and $F(\theta)= \sum F_l e^{il\theta}$, $F_l\in {\bf R}^{2n}$, then in terms of the standard basis for ${\cal H}$, $M_\gamma$ is a ${\bf Z}\times {\bf Z}$ matrix, with each entry $M_{pq}$ being a $2n\times 2n$ matrix, given by $M_{pq}=\gamma _{p-q}$. It can be shown (\cite{Pressley:1988qk})
that this action is in $O_{res}({\cal H})$, so that $LO_{2n}$ is a subgroup 
of $O_{res}({\cal H})$. 

Another fact relating $LO_{2n}$ and $\sss$ is their
number of connected components: it is a property of ${\cal S}$ that 
\be  dim(V^-_\Phi \cap {\cal H}_+) = dim(V^+_\Phi \cap {\cal H}_-) ~ ,\ee
i.e. $I_+(\Phi )=I_-(\Phi)=I(\Phi )$ (the definitions of $V^\pm _\Phi$, 
$I_\pm (\Phi)$, and $J(\Phi )$ are analogous to the ones corresponding 
to $U_{res}$ given in Section \ref{sun}). From this follows that $J(\Phi )=0$ always; 
however, another property of ${\cal S}$  is that smoothly varying
$\Phi $ may change $I(\Phi)$ only by multiples of  2. This means that
 the parity of the dimension $I(\Phi)$ is a topological invariant;
it divides ${\cal S}$ into
two connected components, namely $I(\Phi)$ even and $I(\Phi)$ odd. 
This can be written as follows:
\be \label{Xi} \Xi(\Phi)\equiv I(\Phi) \mbox{ mod } 2\; = -\frac{1}{2} 
tr(M) | _{{\cal H}_+} \; 
\mbox{mod} \; 2 ~ ,\ee
so $\Xi (\Phi )$ is either 0 or 1, and it tells us explicitly 
to which connected
component $\Phi \in {\cal S}$ belongs. So $LO_{2n}$ and ${\cal  S}$ both have
two connected components. We shall use this in section \ref{bnnc}.

Now we construct the symplectic form $\omega $ on $LO_{2n}$ in terms of 
the Killing 
form of $\mathfrak{o} (2n)$, the Lie algebra of $O(2n)$.
We 
begin by defining the Lie algebra $L\mathfrak{o} _{2n}$ associated with 
$LO_{2n}$.
It is the set of smooth maps from the circle to $\mathfrak{o} (2n)$:
\be L \mathfrak{o} _{2n} =\{ \eta :S^1\longrightarrow \mathfrak{o} (2n) \} ~ ,\ee

\no where the commutator is defined by pointwise commutators in $\mathfrak{o} (2n)$. 
The Killing form on $\mathfrak{o} (2n)$ - which is a symmetric, nondegenerate, invariant 
bilinear form - is defined by 
\be K(X, Y)= Tr(ad X \, ad Y) ~ ,\ee

\no where $X$, $Y$ $\in \mathfrak{o}(2n)$, $Tr$ denotes the trace of a 
matrix, and $ad X$ denotes the operator in the adjoint representation of 
$\mathfrak{o}(2n)$ corresponding to $X$. The calculation of the 
Killing form on $\mathfrak{o}(2n)$ is straightforward and gives 
\be K(X,Y)=(2n-2)Tr(XY) ~ .\ee
Now, a symplectic form on $LO_{2n}$ is a two-form, namely a map which 
takes two vector fields and gives a real number. Since a Lie algebra 
in fact consists of vector fields at the identity of the group 
manifold, we can define a symplectic 
form on the identity component $\Omega O_{2n}$ of the loop group as follows 
(see \cite{Pressley:1988qk}): 
\be \label{w} \omega (\eta , \xi)=\frac{1}{2\pi}\int _0^{2\pi} 
K(\eta (\theta), 
\xi ' (\theta)) d\theta ~ . \ee
The form $\omega (\eta , \xi)$ is a map from $L\mathfrak{o}_{2n}\otimes L\mathfrak{o}_{2n}$ 
to the real numbers, and $\eta(\theta)$ and $\xi(\theta)$ are elements 
of $L\mathfrak{o}_{2n}$.
The properties of the Killing form, as well as the decomposition of 
$\eta (\theta)$ and $\xi (\theta)$ into Fourier modes, ensure
that $\omega$ satisfies the 
conditions of a symplectic form, i.e. it is an antisymmetric, 
nondegenerate, closed two-form. Note that this equation
defines a symplectic form for the loop group of any Lie group, with $K$ 
standing for the Killing form of the corresponding Lie algebra.

The Hamiltonian on ${\cal S}$ can be taken to be the same one as equation 
(\ref{hami}),  and it is also a Hamiltonian on $LO_{2n}$ via the embedding 
$LO_{2n}\subset O_{res}({\cal H})$ (equation (\ref{corresp})): we replace $M(x,y)$ by $M_\gamma (x,y)$ which is the integral kernel of $M_\gamma (x)$ defined the same way as in (\ref{kernel}). To get equations of motion for mesons, we would normally
need to derive the Poisson Brackets corresponding to 
the symplectic form given in equation (\ref{w}).
However, we argue below that we can simply rely
on the result for the $U(N)$ case, i.e. that mesons 
on $Gr_1$ satisfy 't Hooft's integral equation, and deduce that  
on the space ${\cal S}$ the equation for mesons will again be 't Hooft's 
integral equation, as is also true for QCD with an $O(M)$ gauge group, to which
we claim that the theory on ${\cal S}$ is equivalent: 

The relationship between $LO_{2n}$ and $\sss$, i.e. the fact that $LO_{2n}\subset
O_{res}({\cal H})$ where $\sss$ is the coset space $\ores / [O({\cal H}_+)
\times O({\cal H}_-)]$ (equation (\ref{sores})), has an analog
in the unitary case: it is true as well (see \cite{Pressley:1988qk}) that $LU_n$, the loop group of
the unitary group, is a subgroup of $\ures$, where we remember that
$Gr_1$ is the coset space $\ures / [U({\cal H}_+) \times U({\cal H}_-)]$ 
(equation (\ref{gr12})). The 
symplectic form on $Gr_1$ given in equation (\ref{sp}) is invariant under the unitary
group action on $Gr_1$ 
and is the unique such form (see \cite{Rajeev:1994tr}); similarly, 
the form in equation (\ref{w}), when it is taken on $LU_n$, is invariant under
the corresponding action in $LU_n$ (in fact, it is invariant under any 
translations in the group). 
Therefore, the meson equation derived on $LU_n$ using equation (\ref{w})
 and the meson equation obtained on $Gr_1$ using equation 
(\ref{sp}) appear to be the same (or closely related), i.e they both are the 
't Hooft integral equation. Similarly, equations obtained directly on $\sss$ using 
a symplectic form invariant under the orthogonal group action, and the
equations derived from equation (\ref{w}) considered on $LO_{2n}$ appear to be the same (or closely related). Since
$\ures$ and $\ores$ are analogous and the embeddings $LU_n \subset \ures$ and 
$LO_{2n} \subset \ores$ have the same structure,  the resulting
meson equations on $\sss$ and on $LO_{2n}$ are analogs of those on $Gr_1$ and $LU_n$, 
and we conclude that we get 't Hooft's equations on $\sss$ (and on $LO_{2n}$) as we did on $Gr_1$.

Now that we established that our theory is consistent with 2DQCD through properties of mesons, it makes sense to consider the topological
invariant of the space ${\cal S}$, which is analogous to the virtual rank of 
$Gr_1$, to be related to baryon number, i.e. to consider baryons to be solitons,
 and we do so below.


\section{Baryon Number Non-Conservation}\label{bnnc}

 \renewcommand{\baselinestretch}{1.4}
\normalsize

As we have established, our phase space has two connected 
components (see equations (\ref{fund}) and (\ref{Xi})), and it is this topological property which we propose here 
to be related
to the baryon number $B$. We will show in this section that the 
properties of 
baryon number given by our theory are consistent with QCD in two dimensions with an odd orthogonal gauge group.

\renewcommand{\baselinestretch}{1.5}
\normalsize

We propose that in the space ${\cal S}$, the topological invariant $\Xi (\Phi)$ given in equation 
(\ref{Xi}) corresponds to the parity of baryon number, i.e. states $\Phi $ with 
$\Xi (\Phi)=0$ have even baryon number, and states with $\Xi (\Phi)=1$
have odd baryon number. Correspondingly, in the space $LO_{2n}$, shrinkable loops correspond to even baryon number, and non-shrinkable loops to odd baryon number. The physical meaning of relating a quantity
 to a topological invariant is that the quantity is conserved. So in this
case, the quantity $Q_B$ defined by

\vskip -.2cm

\be \label{qb} Q_B = B \; \mbox{mod } 2 ~,\ee

\no where $B$ is the baryon number, is conserved, i.e. $\triangle Q_B= 0$, which means 
that baryon number can change 
by multiples of two. In other words, baryon number is conserved only 
modulo two. Physically, this means that baryon-baryon pairs can 
annihilate.\footnote{The idea that baryon number would be conserved only 
mod 2 appeared in a different context in 
\cite{Schafer:1998ef}. }

The equivalence of the theory on ${\cal S}$ to QCD in two dimensions with an orthogonal gauge group is supported
by the following argument: first, baryon-baryon annihilation can be shown directly to occur 
in $O(M)$ QCD, where $M$ here may be even or odd \cite{Witten:1983tx}: unlike $SU(N)$, whose fundamental $N$-dimensional 
representation is complex, $O(M)$ has a real fundamental representation, 
which is the same as its dual. Since quarks are described by the 
fundamental representation while antiquarks are described by the dual 
of that representation, this implies that QCD with an $O(M)$ 
gauge group does not distinguish quarks from antiquarks. Therefore, it 
does not distinguish baryons from antibaryons either, and just as a 
baryon-antibaryon pair can annihilate, so can a baryon-baryon 
pair. This is the same result we 
arrived at above!

The question of baryon number conservation is a bit more subtle; baryon-baryon annihilation does not necessarily imply the conservation of baryon number modulo two: it means only that baryon number can be conserved at most modulo 2, i.e. single baryons might still appear or disappear, in which case there would be no conservation at all. We now show that baryon conservation is determined by the parity of $M$, i.e. it depends on whether the gauge group $O(M)$ is an even orthogonal group, $M=2N$, or an odd orthogonal group, $M=2N+1$. 

The lagrangian of QCD with an orthogonal gauge group has the same form as the one given in equation (\ref{L}), except that the gauge fields $A_\mu $ are now the generators of the orthogonal rather than the unitary group (they are real anti-symmetric rather than complex anti-hermitian) and the quark field $\psi $ is real. It is clear that the lagrangian is invariant under the transformation $\psi \rightarrow -\psi $, the only phase transformation allowed for real fields. This means that the only expectation values which can be non-zero are those which contain an even number of quarks, so that quarks are conserved modulo two. 

We can now see the difference between the even and odd cases.  To form a baryon, we must take a product of $M$ quarks. When $M$ is even, the baryon is the product of an even number of quarks and therefore does not change under $\psi \rightarrow -\psi $, which means that baryons are not conserved at all: Green's functions containing an odd number of baryons can be non-zero, and processes such as $2B\rightarrow 3B$, which both begin and end with an even number of quarks,  may occur. However, the situation when $M$ is odd is different: each baryon consists of the product of an odd number of quarks so the baryon is odd under $\psi \rightarrow -\psi $. Therefore, non-zero correlation functions must necessarily contain an even number of baryons, which means that baryons are indeed conserved modulo 2, consistent with the theory on ${\cal S}$. 

\section{Relation to master fields} \label{matrix}

A classical field, such as the field $M(x,y)$ which we have discussed, representing the large $N$ limit of QCD is also known as a $master \; field$. There is another approach  to master fields 
given in \cite{Gopakumar:1994iq, Douglas:1994kw}, where the mathematical formalism of non-commutative probability theory is used within a matrix model. A master field is in general a classical configuration such that in the large $N$ limit, the values of gauge-invariant Green's functions are given simply by their value at the master field - i.e., no functional integral needs to be done. In \cite{Gopakumar:1994iq, Douglas:1994kw}, the Hilbert space on which the master fields act is given by states generated from the vacuum by creation operators which satisfy the Cuntz algebra, i.e.
\[ a(x)|\Omega >=0, \hskip 1cm a(x)a^{\dagger}(y)=\delta (x-y), \] 
with no further relations. Master fields are given in terms of these creation and annihilation operators, i.e. $M=M(a, a^{\dagger})$. For the case of matrix fields which have independent distributions and are decoupled, there is the form 
\be \label{master1} M(a,a^{\dagger},x)=a(x) + \sum _n M_{n+1} a^{\dagger n}(x) \ee 
(the $M_{n+1}$ turn out to be connected Green's functions). For the more general case in which the matrices are coupled, the form for the master field becomes
\be \label{master2} M(x)=a(x)+\sum _{k=1} ^\infty \psi _{x, y_1, \ldots , y_k}a^{\dagger}(y_1) \cdots a^{\dagger}(y_k) ~ .
\ee 

 \renewcommand{\baselinestretch}{1.6}
\normalsize

An interesting problem would be to make a connection between this framework and the master fields $M(x,y)$ on the Grassmannian manifold described in Sections \ref{sun} and \ref{on}, i.e. to write $M(x,y)$ in terms of these creation and annihilation operators. A construction of the master field for 2DQCD is suggested in \cite{Gopakumar:1994iq}, but we propose a different form for $M(a, a^\dagger)$:
\be \label{masterme} M(x,y)= a^{\dagger}(x)a(y) ~ .\ee
 \renewcommand{\baselinestretch}{1.5}
\normalsize
This form is motivated by the intriguing fact that its commutation relations are identical to 
 equation (\ref{pbM}), except for the $\epsilon$ term which does not have an analog here, and a sign:
\be [M(x,y),M(z,w)]=\delta (y-z)M(x,w)-\delta (w-x)M(z,y) ~. \ee
  Therefore, it appears plausible that a matrix model with this master field could very well provide further insight into 2DQCD.  A full investigation of this possibility is beyond the scope of this paper.

\section{Open Problems} \label{next}
We have extended Rajeev's work to  construct a classical dynamical system which is equivalent  to the large $N$ limit of two dimensional QCD with an $O(2N+1)$ gauge group. We argued that the same equation of motion for mesons would result for the theory
on ${\cal S}$ as for the theory on $Gr_1$, which furthermore is the same 
equation derived within
2DQCD with $O(2N+1)$ and $U(N)$ gauge groups. Considering baryons as topological solitons in our theory, we showed that  baryon number is conserved only modulo 2. We also showed that our model is related to the master field approach to matrix models by providing an explicit formula for the master field that has the same commutation relations as in our model.

There are several directions for further investigations which may now be pursued. We suggest a few below:

It would be interesting to
see what would happen when the symplectic groups or the exceptional
groups serve as the QCD gauge groups. For $Sp(2N)$, we can see immediately that a direct analog of our work would need to be modified. Suppose we propose  an equivalence between a classical dynamical system on an infinite-dimensional Grassmannian manifold, or   the loop group of the symplectic group,  $LSp_{2N}$, and the large $N$ 
limit of 2DQCD with $Sp(2N)$ gauge group. Since $\pi _1 (Sp(2N))={\bf Z}\,$, our construction  would lead to 
conservation of baryon number; however, 
baryons should not even exist in $Sp(2N)$ QCD (they decay into 
mesons; see \cite{Witten:1983tx}).

Another direction one might pursue is quantizing the classical dynamical
system we have described, i.e. promoting the classical 
observables, which are just functions on the phase space, to operators. 
Commutation relations would be given by the Poisson Brackets. This
has been done for the unitary case in \cite{Rajeev:1994tr}, resulting
in a theory of QCD at finite $N$ with a correspondence between $\hbar $ and $1/N$.

In addition, one may calculate the mass for a baryon by minimizing the
Hamiltonian subject to the constraint of odd baryon number. One would then interpret
the resulting minimal energy configuration of odd baryon number to correspond to a 
stable state of one baryon.

Another interesting investigation would be to study chiral symmetry breaking in this system and its relation to the non-conservation of baryon number. See \cite{Auzzi:2008hu} for a recent investigation of chiral symmetry breaking and stability of topological solitons in $SU(N)$ and $SO(N)$ Yang-Mills theories. 

Finally, a full-fledged investigation of the master field approach using the proposed master field (equation (\ref{masterme})) has the potential to shed new light on 2DQCD.

\vskip 1cm
\no {\bf Acknowledgements.} This work was completed under the supervision of Professor Alexander M. Polyakov. It was submitted as part of the requirements for the PhD at Princeton University (preprint number PUPT-2064). The author is very grateful to Professor Polyakov for his inspiration and beauty of thought. She is also grateful to Professors Curtis Callan, Demetrios Christodoulou, John Mather, and Shiraz Minwalla for helpful discussions. This work was funded by the NSF.


\normalsize

\begin{thebibliography}{99}



\bibitem{Rajeev:1994tr}
  S.~G.~Rajeev,
  ``Quantum hadrodynamics in two-dimensions,''
  Int.\ J.\ Mod.\ Phys.\  A {\bf 9}, 5583 (1994)
  [arXiv:hep-th/9401115].



\bibitem{'tHooft:1974hx}
  G.~'t Hooft,
  ``A Two-Dimensional Model For Mesons,''
  Nucl.\ Phys.\  B {\bf 75}, 461 (1974).

\bibitem{Toprak:2002kx}
  E.~Toprak and O.~T.~Turgut,
  ``Large N limit of SO(N) scalar gauge theory,''
  J.\ Math.\ Phys.\  {\bf 43}, 1340 (2002)
  [arXiv:hep-th/0201193].
  
\bibitem{Toprak:2002kw}
  E.~Toprak and O.~T.~Turgut,
  ``Large N limit of SO(N) gauge theory of fermions and bosons,''
  J.\ Math.\ Phys.\  {\bf 43}, 3074 (2002)
  [arXiv:hep-th/0201192].


\bibitem{'tHooft:1973jz}
  G.~'t Hooft,
  ``A Planar Digram Theory for Strong Interactions,''
  Nucl.\ Phys.\  B {\bf 72}, 461 (1974).


\bibitem{Witten:1979kh}
  E.~Witten,
  ``Baryons in the 1/N Expansion,''
  Nucl.\ Phys.\  B {\bf 160}, 57 (1979).


  
\bibitem{Pressley:1988qk} A. Pressley and G. Segal, {\it Loop Groups} (Clarendon, 
Oxford, 1986).


\bibitem{Schafer:1998ef}
  T.~Schafer and F.~Wilczek,
  ``Continuity of quark and hadron matter,''
  Phys.\ Rev.\ Lett.\  {\bf 82}, 3956 (1999)
  [arXiv:hep-ph/9811473].
 
\bibitem{Witten:1983tx}
  E.~Witten,
  ``Current Algebra, Baryons, and Quark Confinement,''
  Nucl.\ Phys.\  B {\bf 223}, 433 (1983).

\bibitem{Gopakumar:1994iq}
  R.~Gopakumar and D.~J.~Gross,
  ``Mastering the Master Field,''
  Nucl.\ Phys.\  B {\bf 451}, 379 (1995)
  [arXiv:hep-th/9411021].



\bibitem{Douglas:1994kw}
  M.~R.~Douglas,
  ``Stochastic Master Fields,''
  Phys.\ Lett.\  B {\bf 344}, 117 (1995)
  [arXiv:hep-th/9411025].

\bibitem{Auzzi:2008hu}
  R.~Auzzi, S.~Bolognesi and M.~Shifman,
  ``Skyrmions in Yang--Mills Theories with Massless Adjoint Quarks,''
  Phys.\ Rev.\  D {\bf 77}, 125029 (2008)
  [arXiv:0804.0229 [hep-th]].


\end{thebibliography}
\end{document}